# Investigating the Best Radio Propagation Model for 4G - WiMAX Networks Deployment in 2530MHz Band in Sub-Saharan Africa

Awal Halifa
Dep't of Electrical Engineering
Kwame Nkrumah Univ. of Science and Technology

E. T. Tchao
Dep't of Computer Engineering
Kwame Nkrumah Univ. of Science and Technology

J. J. Kponyo
Dep't of Electrical Engineering
Kwame Nkrumah Univ. of Science and Technology

## ABSTRACT
One of the salient factors to look at during wireless network planning is developing a modified path loss prediction models to suit a new environment other than the one it was originally designed for. This helps to give accurate predictive outcomes. This paper seeks to demonstrate the effects of applying correction factors on radio propagation model used in planning for 4G-WiMAX network through a comparative analysis between estimated and field data collected on received power for a 4G-WiMAX site. Four existing models were considered for this research; COST 231 Hata, Extended COST 231 Hata, SUI (Stanford University Interim) and Ericsson models. In order to optimize and validate the effectiveness of the proposed models, the mean square error (MSE) and correlation co-efficient were calculated for each model between the predicted and the measured received power for the selected area before and after applying an appropriate correction factor. Based on this, the Extended COST-231 Hata prediction model proved to correlate well with the measured values since it showed least Mean Square Error (MSE) but with highest correlation co-efficient. Through comparative analysis of the corrected models, the Extended COST-231 Hata model could be applied for effective planning of the radio systems in Ghana and the sub-region at large.

## 1. INTRODUCTION
In planning a wireless communication network, it is very important to consider the predictive tool for a signal loss. Both predicted and measurement-based propagation tools reveals that the average RSS decrease logarithmically with respect to distance be it indoor and outdoor wireless channels. This is important when estimating the interference, frequency assignments and evaluation of cell parameters. These are grouped into three: theoretical, empirical and physical [1]. In reality, it is difficult coming out with an accurate prediction model. Practically, researchers that adopt simulation approach apply empirical models, which depend on fitting curves which recreates series of measurement values. However, the validity of an empirical model at a transmission frequency or terrains rather than the one originally used in deriving that model can be established through extensive field measurements taken on a live network. As a result, selecting the best model for a specific geographical terrain becomes extremely difficult due to variations in the land-scopes from terrain to terrain. The validity of the commonly used propagation models therefore becomes ineffective if they are applied on a terrain rather than the one originally used in deriving such models. Several studies conducted in Ghana and some tropical regions have revealed that a lot of the widely used path loss models have lower efficiency relative to the field measured values [2]. Hence, this makes it necessary to investigate most appropriate models that best fits the Ghanaian geographical conditions. The main objective of this research is to undertake a performance comparison between simulation results and field experimental data using the mean square error and correlation co-efficient analysis before and after applying a correction factor to propagation models used in planning a deployed 4G-WiMAX network in the urban centers of Ghana.

## 2. REVIEW OF RELATED LITERATURE
Various scholars have made tremendous contributions in line with path loss model. Zamanillo and Cobo [3] derived a path loss model for UHF spectrum 4 and 5. It was realized in their research that for VHF and UHF band, path loss is independent of frequency, modulation scheme and bandwidth. It was realized that statistical characteristics of the propagation channel in the VHF and UHF spectrum could also be characterized by adopting the model with measurements obtained on a specified frequency irrespective of the frequency under consideration. Hanchinal [4] also conducted a study by making a comparative analysis between the estimated results of wireless models and field measurement data. His conclusions were that the COST- 231 and SUI models give the most suitable predictions for the various terrain categories specifically for the urban and suburban terrains. The author further came out with a more corrected model for estimating the path loss in urban terrains. As a result of the findings in [4], the corrected path loss model was derived by comparing between the estimated path loss values and field measurement values.

Abraham, et al [5] also conducted a study which centered on comparing the path loss prediction model with field values for the macro cellular terrain. The findings revealed that, Hata model gave more accurate and precise path loss predictions for the macro cellular terrain. Their model produced an MSE of 2.37 dB which was far less than the minimum acceptable MSE of 6 dB for a good signal propagation.

All these propagation models as explained in the literature survey were obtained from the measurement data taken under European conditions and Asian terrains which aren't similar to the peculiar conditions of the Sub-Saharan Environment. It is possible to adopt the same or similar models for correcting pathloss models used in planning networks in sub-region by applying correction factors to these models. Once these terrain related parameters could be calculated and additional field measurements taken for validation, the necessary correction terms could be appropriately applied as required to improve the accuracy in using these models to predict the received signal power.

## Description of the Measurement Terrain

European climatic condition is classified as been precipitated or humid unlike the sub-Saharan region which has less precipitation or completely dry weather conditions which can affect the performance of wireless propagation models leading to poor quality of service. In addition, in the sub-Saharan countries, building structures are not properly sited unlike in the tropical countries where there are stringent laws and regulation [6]. In Ghana, it is possible for an operator to site its Base Transceiver Station to have a clear line of site (LoS) with neighboring cell sites only for the LoS to be obstructed by the erection of an unapproved structure which wasn't factored during the planning phase of the network systems. This makes network planning very difficult which leads to poor quality of service for subscribers.

In order to determine how the peculiar terrain affects networks deployment, field measurement was carried around the University of Ghana Campus. This area provides a measure of the WiMAX network's radio distance in a typical urban area. The distribution of Customer Premise equipment in the study area is shown in Figure 1. The simulation model adopted for analyzing the capacity and coverage range of the deployed WiMAX network is based on the behavior of users and the distribution of CPEs in the cell. The simulation was done using the stochastic distribution of CPEs within the cell site as shown in Figure 2. The model adopted the approach used in [7] which used circular placement of nodes in a hexagon with one WiMAX Base Station and Subscriber Stations (SS) which were spaced apart from the Base Station (BS). The BS are fixed and mobile nodes since mobility has been configured.

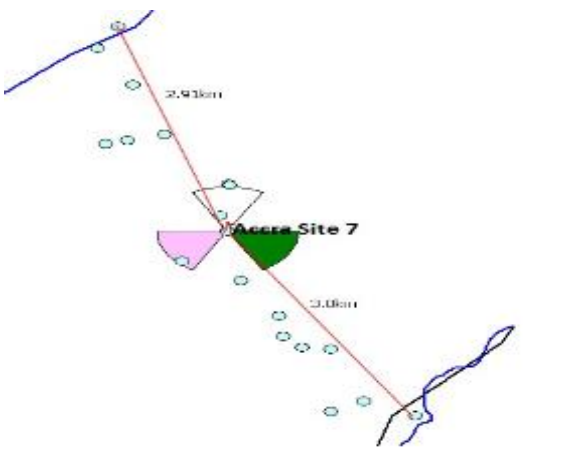

**Figure 1: Distribution of CPE in study area**

The measurement area was selected due to its urban features suitable for deploying outdoor wireless networks. The measurement set-up was composed of a GPS, dongle XCAL-X, a laptop with a XCAL-X software, WiMAX PCMCIA CARD and Programmable field strength analyzer. The measurements were divided into RSS and throughput. The experimental set-up as shown in Figure 2 was used in testing real live performance of the WiMAX network under Sub-Saharan African condition.

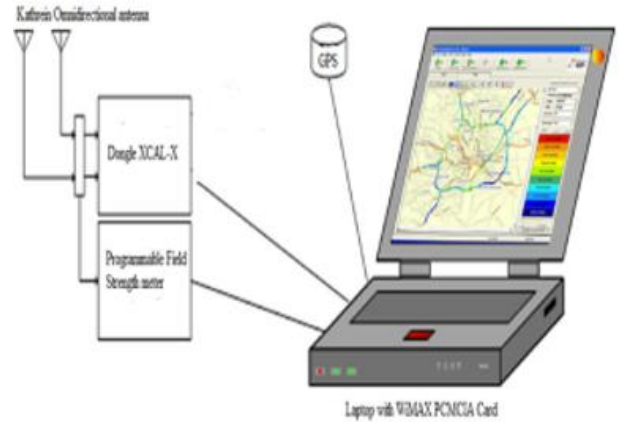

**Figure 2: Field trial experimental set-up**

Several locations within the measurement areas were found to be fairly obstructed, with a few sections covered by dense foliage. The RSS test taken in over 16,000 locations within the cell site. The results of the drive test values have been summarized in Figure 3.

The highest measured value of -45 dBm was recorded at the cell centre at about 500m away from the BTS. This value was fairly good when compared with the simulated cell centre RSS value of -40dBm. The measured cell edge RSS value was found to be -100dBm at 4km.

.

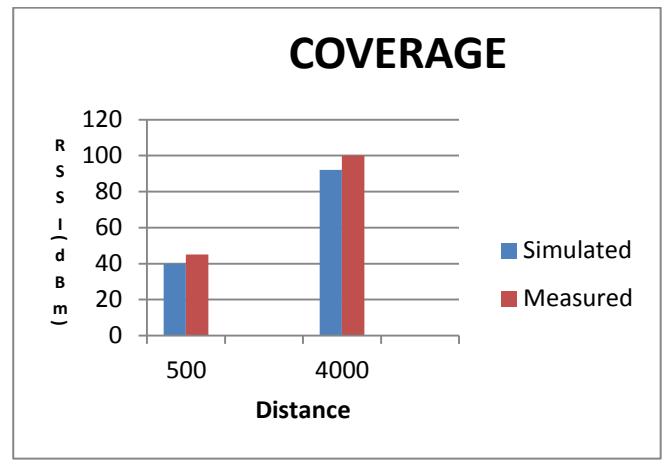

**Figure 3: Coverage Comparison (-dBm)**

The results of the throughput measurement tests have been summarized in Figure 4. The maximum and minimum field measured downlink throughputs were 6.1 Mbps and 300kbps respectively. These values show a large variation when compared with the simulated maximum and minimum downlink throughput values of 8.82 Mbps and 1.2 Mbps respectively.

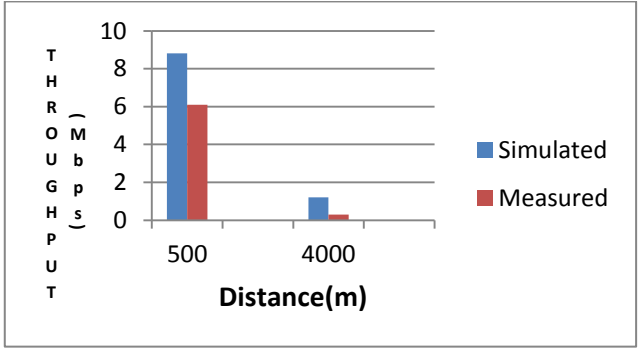

**Figure 4: Throughput Comparison**

Analysis of these field measurement results seem to support the assumption made in [2] and [8] that the correction factor for pathloss models which have not been specified for the Sub-Saharan African environment could have contributed to the differences between the simulated throughput per sector and the measured values. Hence there is the need to study how propagation pathloss affects network deployment in the study environment.

## 3. PATH LOSS MODELS EMPERICAL PROPAGATION MODELS

Generally, the real terrain becomes extremely difficult to model accurately. Practically, several simulation findings adopt empirical models, which are derived, based on field data from different actual terrains on a live network [9]. This section describes a number of commonly used empirical models.

### Free Space Path loss Models

This model explains the phenomenon where there is a clear visible transmission path between the transmitting and the receiving antennas. It is directly proportional to the square of the distance between the receiving and transmitting antennas, and to the square of the frequency of the wireless signal as well [10]. The Free Space Path loss is estimated using the equation [11][12]:

$$PL_{FSP}(dB) = 32.45 - 10\log_{10} G_t + 20\log_{10} f + 20\log_{10} d \qquad (1)$$

Where:

$P_r$ = Receive (rx) power

$P_t$ = transmit (tx) power

$G_t$ = transmit gain

$G_r$ = Receiver gain

d = antenna separation distance

$P_{FSP}$ = free space path loss

### COST 231 Hata Path Loss Model.

COST-231 model is the modified form of Okumura-Hata model for frequencies above 1.8GHz [13]. It has been found to be appropriate for planning wireless networks in medium and large cities having a base station antenna height higher than the surrounding structures. The urban terrain path loss can be estimated using [14]:

$$PL_{50}(urban) = 46.3 + 33.9*\log_{10} f_c - 13.82*\log_{10} h_{te} - (44.9 - 6.55\log_{10} h_{re})*\log_{10} d + C_m \qquad (2)$$

For;

$C_m$ = 0 dB; medium suburban whiles for Metropolitan it is 3 dB

$f_c$ = frequency band from 150-2000 MHz

$h_{te}$ = effective tx antenna height within 10-200m range

$h_{re}$ = effective rx antenna height within 1-10m range

### Extended COST 231 Hata Model

Okumura Hata model is mostly applicable in empirical propagation model, which is based on the Okumura model [15]. This model is derived for the UHF spectrum. Initial recommendations of ITU-R P.529 revealed that model was limited to 3500MHz. The Extended COST 231 is modeled as [16]:

$$PL = A_{fs} + A_{bm} - G_b - G_r \qquad (3)$$

For;

$A_{fs}(dB)$ = free space attenuation

$A_{bm}(dB)$ = Basic median path loss

$G_b$ = transmit antenna height gain factor

$G_r$ = rx antenna gain factor

### Stanford University Interim (SUI) Model

Acceptable standards for the spectrum less than 11 GHz have channel models derived by Stanford University called SUI models [13]. The SUI models are grouped into: A, B and C terrains as shown in Table 1. Terrain A has highest path loss and is suitable for hilly. They are applied to the 3500 MHz spectrum that is in used in some Countries. Terrain B is used in areas that are usually flat having a moderate to heavy tree density or hilly having a light tree density. The SUI model estimates path loss using: for;

$$PL_{SUI} = A + 10\gamma \log_{10}\left(d/d_0\right) + X_f + X_h + s \qquad (4)$$

for $d > d_0$

d: BS - receiving antenna distance[m]

$d_0$ : Reference distance, [100m]

$X_f$ Correction factor for frequencies beyond 2 GHz

$X_h$ : Correction factor for receiver antenna height (m)

s: Correction for shadowing in dB and

$\gamma$ : Path loss exponent.

The path loss exponent γ and standard deviation 's' are selected through statistical analysis. Log-normally distributed factor 's' denotes shadow fading due to environmental clutters on the transmission path having values within 8.2 dB - 10.6 dB range [8].

**Table-1: The parameter figures for different terrain of SUI model**

| Model Parameter | Terrain A | Terrain B | Terrain C |
|---|---|---|---|
| A | 4.6 | 4 | 3.6 |
| $b(m^{-1})$ | 0.0075 | 0.0065 | 0.005 |
| c(m) | 12.6 | 17.1 | 20 |

The frequency correction factor and the correction factor for receiver antenna height for the models are:

$$X_f = 6.0 \log_{10}\left(\frac{f}{2000}\right)$$

For Terrains A & B:

$$X_h = -10.8 \log_{10}\left(\frac{h_r}{2000}\right)$$

For Terrain C;

$$X_h = -20 \log_{10}\left(\frac{h_r}{2000}\right)$$

Where:

f = frequency (MHz)

$h_r$ is receiver antenna height (meters)

These stated correction factors for this model is applied for estimating path loss for rural, urban and suburban terrains.

### Ericsson Model
In predicting path loss, the network operators sometimes adopt a path loss estimating tool engineered by Ericsson Company which is termed as the Ericsson model. This model depends on the corrected Okumura-Hata model to conform to variation in parameters relative to the propagation terrain [4]. Path loss estimated by Ericsson model is given as;

$$PL = a_0 + a_1 \log_{10}(d) + a_2 \log_{10}(h_b) + a_3 \log_{10}(h_b)$$
$$* \log_{10}(d) - 3.2\left[\log_{10}(11.75 h_r)^2\right] + g(f) \quad (5)$$

Where: $g(f) = 44.49 \log_{10}(f) - 4.78\left[\log_{10}(f)\right]^2$

Parameters are;
f: freq.(MHz)

$h_b$ : transmit antenna height (m)

$h_r$ : receive antenna height (m)

## 4. METHODOLOGY
The methodology implemented in this research is summarized in Figure 5. With this procedure, one can implement it on a new environment where the use of correction factor might be required.

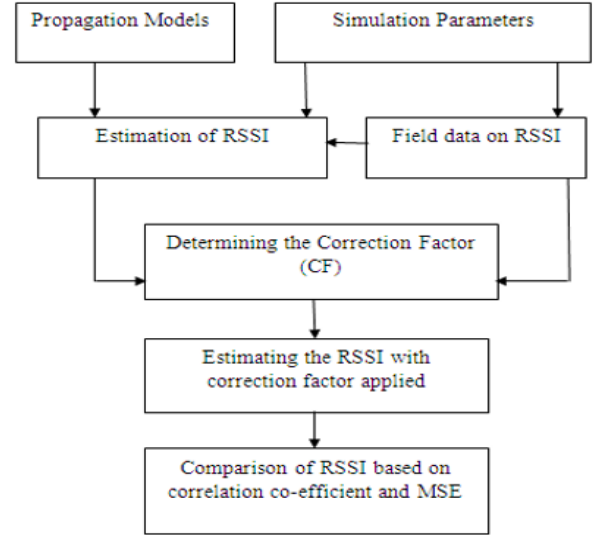

**Figure 5: Research Methodology**

This research methodology summarizes the procedure used in determining an appropriate correction factor for the selected signal prediction model using field data. The received signal power which is evaluated based on the modified signal prediction tool is correlated with the received signal power obtained from the field data. In this adopted research methodology, the propagation pathloss estimation with the appropriate simulation parameters and the field data on received signal strength would be collected. After collecting data on RSS, analysis would be done with the MSE to determine appropriate factors of correction. The difference between the measured and the estimated RSS with the modified model will be analyzed. The difference in values between these two is what would be termed as the correction factor. Once these correction factors for the chosen models have been estimated, they are then applied to the models and the received signal powers for each model is then re-estimated. To get the best suited model, RSS performances are compared based on two criteria: correlation co-efficient and mean square error analysis. When this is done the model having the lowest mean square error but highest correlation co-efficient is selected as the best and most suitable model. Hence, these factors would be used to estimate the accurate received signal strength prediction for the Sub-Saharan terrain profile.

### Estimating the RSS and Mean error
The received signal power, which is estimated as a function of the respective distance away from the BS can be calculated using the relation:

$$P_r = P_t + G_t - PL - \eta_{tx} - \chi \quad (6)$$

Where;

$P_r$ =. Receive power

$P_t$ = Transmit power

$G_t$ = Transmit antenna gain (dBi)

$PL$ = estimated path loss (dB)

$\eta_{tx}$ = loss in transmit feeder cable (dB)

$\chi$ = loss due to difference in the T-R antenna polarization

Using the received power expression which compensates for the gain of the receiving antenna gives an accurate prediction of the receive power and hence in this work, we will estimate the received power using;

$$P_r = P_t + G_t + G_r - PL - \eta_{tx} - \chi \qquad (7)$$

After the received power evaluations have been done, a comparative analysis based on the criteria for correlation coefficient, r and mean square error (MSE) will subsequently be done. In optimizing the pathloss models using the relations in (1) – (7), the efficiency of the optimized models could be validated on the basis of the mean square error and correlation co-efficient using the respective equations:

$$MSE = \frac{1}{N} \sum_{i=1}^{N} \left( P_{r,1}^e - P_{r,1}^m \right)^2 \qquad (8)$$

$$r = \frac{\sum P_r^m P_r^e - \frac{\sum P_r^m \sum P_r^e}{N}}{\sqrt{\left( \sum (P_r^m)^2 - \frac{(\sum P_r^m)^2}{N} \right)\left( \sum (P_r^e)^2 - \frac{(\sum P_r^e)^2}{N} \right)}} \qquad (9)$$

Where:

$r$ : :Correlation coefficient,

$P_r^m$ : measured RSS,

$P_r^e$ : estimated RSS

$N$ : Sample size

From the field measured results summarized in Figure 3, it is evident that there exists a significant difference between the field measured data and that of the predicted. As discussed earlier, these differences in measurement errors is what form the basis for undertaking this study to correct the models. After the determination of the mean error parameter, the correction terms could either be added or subtracted from the respective path loss equation to give a modified propagation model which can produce an accurate prediction of signal power with reference to a Base Station.

From the model optimization methodology shown in Figure 6, the simulation will be carried out in the following procedures:

i. Path loss from the BTS transmitter on the University of Ghana Campus will be estimated for distances relative to the measurements for received power obtained for the selected models using the simulation parameters in Table 3.

ii. The received power from the BS transmitter will be evaluated for distances corresponding to the field data for each of the selected models.

iii. Correcting factors for the selected model will be calculated.

iv. Evaluation of RSS of the models having correction factor is subsequently done.

v. Correlation Coefficient and the MSE between the predicted and field data on RSS will be estimated to serve as a basis to modify the respective models.

vi. Comparative analysis between correlation coefficient and the MSE is done.

## 5. RESULTS AND DISCUSSION

This section presents the results from the field measurement study. The preliminary measurements results have been presented in Table 2. From the results in can be seen that there is a wide variation been measured and estimated results. This confirms the primary assumption of this study.

It is a fact that, the major propagation model, Hata model, which forms the basis for developing the four selected models, was developed for some propagation scenarios different from the peculiar Sub-Saharan African terrain. This could be a contributing factor for the significant difference that exists between the estimated received power values and those obtained from measurement on the live network in this study. This is due to the fact that measurements were collected in less precipitation regions as in Sub-Saharan terrain than the original Hata model in tropical regions.

Mean square error and correlation coefficient of the models using real field values from a live network were used as criteria for assessing the efficiency of the models as quantitative measures of accuracy. This finding proved that, all the selected models equally correlate with the field measurement data. But in this sense, the criterion used in selecting the best model is the model that has least errors (lowest MSE) after applying the correction factors and also with highest correlation as indicated in Table 4. However, through comparative analysis based on the calculated figures the Extended COST-231 Hata Model was considered best because it has the lowest Mean Square Error of 6.254 dB.

The results in Table 4 serves as a basis for modifying the extended COST 231 model for the Sub Saharan African environment as:

$$PL_{[\text{modified COST-231 Hata}]} = 46.3 + 33.9 \log_{10}(f) - 13.82 \log_{10}(h_b)$$
$$- ah_m + \left[ 44.9 - 6.55 \log_{10}(h_b) \right] \log_{10} d + c_m - 7.845 \qquad (9a)$$

This equation can be summarized as:

$$PL_{[\text{modified COST-231 Hata}]} = A_{fs} + A_{bm} - G_b - G_r - 7.845 \qquad (9b)$$

The correlation coefficients and the MSE between the field data on received power and the predicted with correction terms applied to selected models with results tabulated in Table 4 accordingly.

Comparing the results of MSE's, given in Table 4, it can be deduced that the new proposed model will be more accurate in predicting the actual path loss. Based on this, a modified Extended COST-231 Hata model for the prediction of path loss for WiMAX networks deployment in the 2500-2530 MHz bandin urban environment of Greater Accra is developed in (9b).

This modified model gives a high degree of accuracy and is able to predict path loss with smaller mean error relative to the original Extended COST 231 Hata model. The modified Extended COST 231 Hata model shows greater performance and higher accuracy than the original Extended COST-231 Hata model based on the mean square error and correlation coefficient analysis.

**Table 2: Field Data on RSS and Predicted RSS**

| Distance(m) | Received signal power(dBm) | | | | |
|---|---|---|---|---|---|
| | | Estimated using | | | |
| | Measured RSSI | COST-231 Hata | Extended COST-231 Hata | SUI | Ericsson |
| 4200 | -73 | -97.94 | -91.87 | -66.86 | -100.58 |
| 4000 | -83 | -97.21 | -91.14 | -65.98 | -99.93 |
| 3900 | -87 | -96.84 | -90.76 | -65.52 | -99.6 |
| 3800 | -92 | -96.45 | -90.37 | -65.05 | -99.26 |
| 3600 | -81 | -95.64 | -89.56 | -64.07 | -98.55 |
| 3600 | -87 | -95.64 | -89.56 | -64.07 | -98.55 |
| 3500 | -79 | -95.22 | -89.14 | -63.56 | -98.17 |
| 3500 | -78 | -95.22 | -89.14 | -63.56 | -98.17 |
| 3500 | -80 | -95.22 | -89.14 | -63.56 | -98.17 |
| 3300 | -77 | -94.34 | -88.27 | -62.5 | -97.4 |
| 3200 | -75 | -93.88 | -87.81 | -61.94 | -96.99 |
| 3000 | -75 | -92.92 | -86.86 | -60.77 | -96.14 |
| 2800 | -73 | -91.89 | -85.86 | -59.52 | -95.23 |
| 2700 | -74 | -91.34 | -85.33 | -58.87 | -94.75 |
| 2700 | -72 | -91.34 | -85.33 | -58.87 | -94.75 |
| 2000 | -71 | -86.86 | -81.06 | -53.43 | -90.8 |
| 1900 | -70 | -86.09 | -80.34 | -52.51 | -90.12 |
| 1800 | -68 | -85.28 | -79.59 | -51.53 | -89.41 |
| 1600 | -67 | -83.52 | -77.97 | -49.39 | -87.85 |
| 1500 | -67 | -82.56 | -77.09 | -48.23 | -87 |
| 1400 | -65 | -81.53 | -76.15 | -46.98 | -86.09 |
| 1300 | -69 | -80.42 | -75.16 | -45.64 | -85.12 |
| 1200 | -70 | -79.22 | -74.1 | -44.19 | -84.06 |
| 1200 | -65 | -79.22 | -74.1 | -44.19 | -84.06 |
| 1100 | -64 | -77.92 | -72.95 | -42.61 | -82.91 |
| 900 | -65 | -74.93 | -70.35 | -38.98 | -80.27 |
| 800 | -63 | -73.17 | -68.86 | -36.85 | -78.71 |
| 800 | -63 | -73.17 | -68.86 | -36.85 | -78.71 |

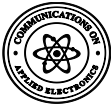

| | | | | | |
|---|---|---|---|---|---|
| 700 | -60 | -71.17 | -67.18 | -34.43 | -76.95 |
| 700 | -62 | -71.17 | -67.18 | -34.43 | -76.95 |
| 600 | -55 | -68.87 | -65.29 | -31.64 | -74.92 |
| 570 | -61 | -68.1 | -64.67 | -30.71 | -74.24 |
| 560 | -57 | -67.84 | -64.45 | -30.39 | -74.01 |
| 550 | -59 | -67.57 | -64.24 | -30.07 | -73.77 |
| 530 | -57 | -67.01 | -63.79 | -29.4 | -73.28 |
| 520 | -52 | -66.73 | -63.56 | -29.05 | -73.03 |
| 500 | -61 | -66.14 | -63.1 | -28.34 | -72.52 |
| 500 | -58 | -66.14 | -63.1 | -28.34 | -72.52 |
| 500 | -62 | -66.14 | -63.1 | -28.34 | -72.52 |
| 470 | -51 | -65.22 | -62.36 | -27.22 | -71.7 |
| 450 | -59 | -64.57 | -61.85 | -26.44 | -71.13 |
| 420 | -62 | -63.54 | -61.05 | -25.19 | -70.22 |
| 420 | -51 | -63.54 | -61.05 | -25.19 | -70.22 |
| 415 | -50 | -63.36 | -60.91 | -24.97 | -70.06 |
| 400 | -61 | -62.81 | -60.48 | -24.3 | -69.57 |

**Table 3: Base Station Simulation Parameters**

| Base Station Parameter | Value |
|---|---|
| Transmit Power | 30dBm |
| Transmitter Gain | 20dBi |
| Receiver Gain | 18dBi |
| Loss in transmitter feeder cable | 1.2dB |
| Loss due to variation in the transmitter receiver antenna polarizations | 3dB |

**Table 4: Propagation Model Performance Before and After Applying Correction Factors**

| Before Applying Correction Factor | | | | After Applying Correction Factor | | |
|---|---|---|---|---|---|---|
| Propagation Models | MSE | Correlation Coefficient, r | Correction Factor (CF) | MSE | Correlation Coefficient, r | Correction Factor (CF) |
| **COST 231 Hata** | 181.9705 | 0.9188 | 12.5301 | 24.952 | 0.9188 | 12.5301 |
| Extended COST 231 Hata | 79.7012 | 0.9217 | 7.8451 | **6.254** | 0.9217 | 18.1554 |
| **SUI** | 547.5657 | 0.9188 | -22.3657 | 47.3428 | 0.9188 | -22.3657 |
| **Ericsson** | 317.2153 | 0.9188 | 17.2884 | 18.3261 | 0.9188 | 17.2884 |

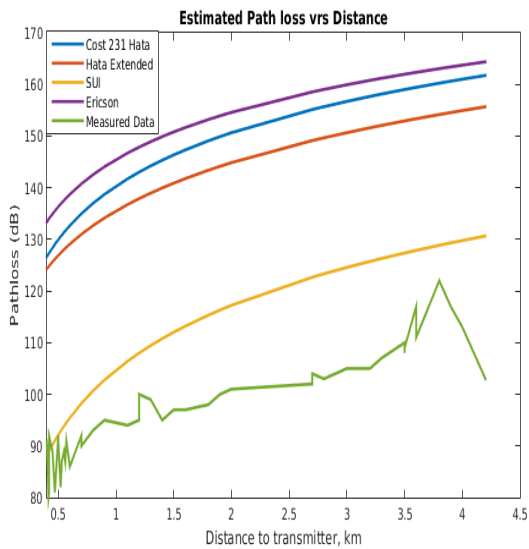

Figure 6: Simulated Pathloss vrs measured Pathloss

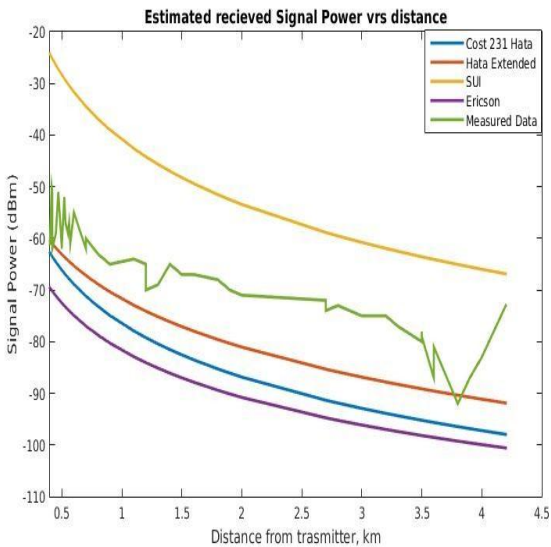

Figure 7: Simulated Received power vrs measured data

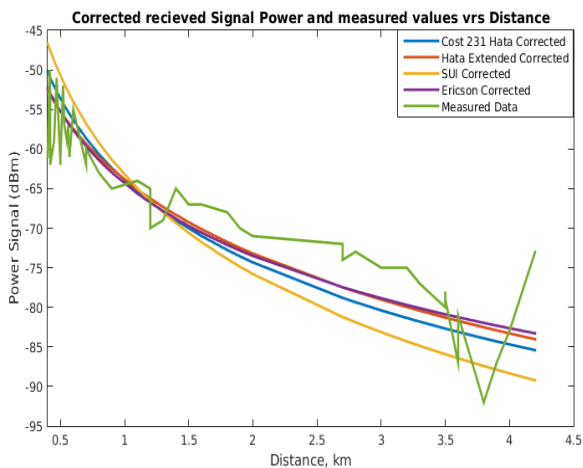

Figure 8: Estimation of Received Power (RSS) with modified models

From the results obtained in Tables 2 and 4, and the optimized model in (*9b*), the pathloss and its estimated Received power simulations have been done and shown in Figures 6 and Figure 7 respectively. It is evident in Figure 8, that the estimated RSS with correction factors applied to the models show that all the models correlates equally with the field values.

## 6. CONCLUSION

This research proves that there is absolutely no specific model which can be used to give consistent results for all propagation environments due to differences in climatic and geographic locations. It was further realized that the characteristics of wireless propagation models differs relative to the frequency of transmission. This research has revealed that Extended COST-231 Hata Model is the best wireless propagation models for WiMAX network deployment in University of Ghana Campus, Accra within the 2500-2530 MHz band. This will help network operators to accurately design future WiMAX network with optimum network throughput with enhanced quality of service to address the ever-growing demand for wireless services in Accra city and the sub-region at large. It is recommended that more research on varying terrain parameters that hinder signal transmission should be considered to give an optimized model that fits well with chosen terrain and analyzing the impact varying frequency bands that proves suitable for chosen terrain.

## 7. REFERENCES

[1] J. Chebil, A. K. Lawas, and M. D. Rafiqul Islam, "Comparison between measured and predicted path loss for mobile communication in Malaysia," *World Appl. Sci. J.*, vol. 21, no. special issue 1, pp. 123–128, 2013.

[2] Eric Tutu Tchao, Kwasi Diawuo and W.K. Ofosu, "On the Comparison Analysis of 4G-WiMAX Base Station in an Urban Sub-Saharan African Environment", Journal of Communication and Computer October, 2013, pp 863-872, ISSN 1548-7709, US.

[3] J. M. Zamanillo and B. Cobo, "Path-Loss Model for UHF Bands IV and V," 8th WSEAS Int. Conf. SIMULATION, Model. Optim., vol. 2, pp. 337–339, 2008.

[4] C. S. Hanchinal, "A Survey on the Atmospheric Effects on Radio Path Loss in Cellular Mobile Communication System," Int. J. Comput. Sci. Technol., vol. 8491, no. 1, pp. 120–124, 2016.

[5] D. C. Abraham, D. D. Danjuma, S. M. Suleiman, and N. D. Academy, "A Discrete Least Squares Approximation Based Algorithm For Empirical Model Adaptation," J. Multidiscip. Eng. Sci. Technol., vol. 3, no. 6, pp. 5116–5122, 2016.

[6] E. T. Tchao, W K Ofosu, K Diawuo: Radio Planning and Field Trial Measurement of a Deployed 4G WiMAX Network in an Urban Sub-Saharan African Environment; International Journal of Interdisciplinary Telecommunications and Networking. September, 2013; 5(5): pp 1-10.

[7] J.D Gadze, L. A. Tetteh and E. T Tchao. "Throughput and Coverage Evaluation of a Deployed WiMAX Network in Ghana", International Journal of Computer Science and Telecommunications, July, 2015. Volume 7, Issue 5, pp 18-26.

[8] E. T. Tchao, W. K. Ofosu, K. Diawuo, E. Affum and Kwame Agyekum "Interference Simulation and Measurements for a Deployed 4G-WiMAX Network in an Urban Sub-Saharan African Environment": International Journal of Computer Applications (0975 - 8887) Volume 71 - No. 14, pp 6-10, June 2013

[9] C. Temaneh-Nyah and J. Nepembe, "Determination of a Suitable Correction Factor to a Radio Propagation Model for Cellular Wireless Network Analysis," 2014 Fifth Int. Conf. Intell. Syst. Model. Simul., vol. 10, no. 35, pp. 175–182, 2014.

[10] C. Paper, "Simulation and Analysis of Path Loss Models for WiMax Communication System," Res. Publ., vol. 3, pp. 692–703, 2015.

[11] D. Alam, S. Chowdhury, and S. Alam, "Performance Evaluation of Different Frequency Bands of WiMAX and Their Selection Procedure," Int. J. Adv. Sci. Technol., vol. 62, pp. 1–18, 2014.

[12] C. Julie, "Site specific measurements and propagation models for GSM in three," Am. J. Sci. Ind. Res., vol. 2, no. 2003, pp. 238–245, 2013.

[13] P. A. Vieira and E. R. Vale, "Field Trial and Analysis of a Received Radio Signal in a 3 . 5 GHz Band Maritime Environment," J. Microwaves, Optoelectron. Electromagn. Appl., vol. 14, no. September, pp. 194–204, 2015.

[14] J. H. Whitteker, "Physical optics and field-strength predictions for wireless systems," in IEEE Journal on Selected Areas in Communications, 2002, pp. 515–522.

[15] D. Alam, S. Chowdhury, and S. Alam, "Performance Evaluation of Different Frequency Bands of WiMAX and Their Selection Procedure," vol. 62, pp. 1–18, 2014.

[16] A. Kale, S and Jadhav, "An Empirically Based Path Loss Models for LTE Advanced Network and Modeling for 4G Wireless Systems at 2.4 GHz, 2.6 GHz and 3.5 GHz," Int. J. Comput. Appl., vol. 7, no. 1, pp. 36–43, 2013.